\pdfoutput=1
\documentclass[11pt]{article}
\usepackage[T1]{fontenc}
\usepackage{lmodern}
\usepackage[margin=1in]{geometry}
\usepackage{amsmath,amssymb}
\usepackage{booktabs}
\usepackage{xcolor}
\usepackage[hidelinks]{hyperref}
\hypersetup{
  pdftitle={The Missing Fourth Term for the Emulation Tensor Memory Equilibrium (TME) Model: The Residue Deconstruction Cost},
  pdfauthor={Harun Bayraktar, John Gunnels, Peter Caday},
}
\newcommand{\Plow}{P_{\mathrm{low}}}
\newcommand{\Pfp}{P_{\mathrm{fp64}}}
\newcommand{\Pint}{P_{\mathrm{int}}}
\newcommand{\Bmem}{B_{\mathrm{mem}}}
\newcommand{\nout}{n_{\mathrm{out}}}
\newcommand{\nin}{n_{\mathrm{in}}}
\newcommand{\OI}{\mathrm{OI}}

\title{\Large\bfseries The Missing Fourth Term for the Emulation\\
  Tensor Memory Equilibrium (TME) Model:\\
  The Residue Deconstruction Cost}
\author{Harun Bayraktar \qquad John Gunnels \qquad Peter Caday\\[4pt]
  {\normalsize NVIDIA Corporation}\\
  {\small\texttt{\{hbayraktar, jgunnels, pcaday\}@nvidia.com}}}
\date{\today}

\begin{document}
\maketitle

\begin{abstract}
The Tensor--Memory Equilibrium (TME) model of \emph{FP8 is All You Need (Part~1)} calculates the
execution time of Ozaki~Scheme~II emulation of fp64 as the maximum of a tensor-core term and a
High-Bandwidth Memory (HBM) traffic term, plus a per-output reconstruction term. However, it omits
the per-\emph{input} deconstruction cost: every streamed fp64 operand must be scaled, rounded, and
reduced modulo each of the $r$ moduli on SIMT pipes before any matrix multiply can
issue. In this note we add this fourth term, calibrate its constant from the cuBLAS emulation path,
and derive a closed-form operational-intensity threshold $\OI^{*} = c_q r \Pfp/(8\Pint)$ below
which emulation cannot match native fp64 regardless of tensor-core throughput. On the NVIDIA B300 GPU the
threshold is $\OI^{*}\approx 0.56$~FLOP/B. As a result, GEMV, SpMV, and low-batch GEMV, which are
the memory-bound kernels the original paper claims to accelerate, are limited to $0.3$--$0.9\times$
of native performance, and the 7-point stencil to $1.8\times$ rather than the claimed $3.1\times$.
Dense GEMM is unaffected as expected. We also show that precomputing and storing the residues moves
the same cost into the bandwidth term, and we state the instruction count that an implementation
would have to achieve to invalidate the bound.
\end{abstract}

\section{Introduction}

Matsuoka~\cite{matsuoka2026fp8} argues that FP8 tensor cores combined with Ozaki
Scheme~II~\cite{ozaki2025scheme2} can replace native fp64 hardware across the HPC kernel spectrum,
including memory-bound kernels, and supports this with the Tensor--Memory Equilibrium (TME)
performance model. The model is a roofline-style bound: the execution time is the maximum of a
tensor-core term and a High-Bandwidth Memory (HBM) traffic term, plus a reconstruction term. In this note we show that the
model omits a cost that is first order for exactly the memory-bound kernels it is used to evaluate,
and we quantify the effect on the kernel-by-kernel speedups reported in the paper.

The analysis here is written against version~3 of the paper~\cite{matsuoka2026fp8}, whose Eq.~(9) does
not include the term. Version~4, posted after this note was shared with the author, adds the term
to the model.

\section{The corrected model}
\label{sec:model}

The TME model of the original paper~\cite{matsuoka2026fp8} calculates the execution time of
emulated execution as its Eq.~(9):
\begin{equation*}
T_{\mathrm{emu}} \;=\; \max\!\left(\ \frac{\alpha W}{\Plow},\ \ \frac{\beta Q}{\Bmem}\ \right) \;+\; \gamma\,\nout .
\end{equation*}
It accounts for the tensor-core work ($\alpha$), the HBM traffic ($\beta$), and the
\emph{output}-side Chinese Remainder Theorem (CRT) reconstruction ($\gamma\,\nout$). It does not
account for the \emph{input}-side deconstruction. Every fp64 operand element must be scaled,
rounded, and reduced modulo each of the $r$ moduli before a single matrix multiply-accumulate (MMA)
instruction can issue. This work is elementwise and nonlinear (round, mod), it has no MMA
formulation, and it therefore runs on the SIMT pipes (integer, fp32, etc., depending on the algorithm and GPU architecture). To simplify the model, we assume a physical SIMT integer pipe is the bottleneck, so that execution time can be modeled in terms of integer instructions.

Let $c_q$ be the number of integer instructions required to convert one element to \emph{one}
modulus, and let $\Pint$ be the sustained SIMT integer instruction rate. Since the streamed input
consists of $Q/8$ fp64 elements, the corrected model is
\begin{equation}
\boxed{\;T_{\mathrm{emu}} \;=\; \max\!\left(\ \frac{\alpha W_{\mathrm{mma}}}{\Plow},\ \ \frac{\beta Q}{\Bmem},\ \
\frac{c_q\, r\, Q}{8\,\Pint}\ \right) \;+\; \gamma\,\nout\;}
\label{eq:corrected}
\end{equation}
The new term is the counterpart of $\gamma\,\nout$: reconstruction is charged per output element
and deconstruction per input element. For memory-bound kernels $\nin = Q/8 \gg \nout$ (for GEMV,
$\nin = W/2$), so the omission in the original model is first-order.

Note that equation~(\ref{eq:corrected}) very optimistically assumes that the preprocessing costs can be completely overlapped with MMA work, assuming $\alpha W_{\text{mma}}/P_{\text{low}} \geq c_qrQ/8P_{\text{int}}$. In reality, there are numerous additional overheads not counted here:
\begin{itemize}
    \item Preprocessing competes for resources (register/cache/memory bandwidth and register/shared memory space, power) with MMA operations.
    \item Some latency necessarily exists between the start of preprocessing and the availability of sufficient preprocessed data to begin MMA operations.
    \item If preprocessed data is shared by multiple MMA units, it must either be written back to cache/memory for usage by other threads, increasing bandwidth usage, or recalculated by each MMA unit, in which case the $c_q$ term in~\eqref{eq:corrected} must be multiplied by a factor $> 1$ that takes into account the redundant calculation. Section~\ref{sec:storing} considers the first approach in more detail.
\end{itemize}

\section{Tile granularity of the MMA term}
\label{sec:tile}

The MMA term must account for the executed work rather than the useful work. The native tile of the
fp64 tensor core is $[8{\times}4]\times[4{\times}8]$ (\texttt{mma.m8n8k4.f64}), and the minimum
accumulator width of the FP8 MMA is likewise $n = 8$ columns (\texttt{mma.m16n8k32} for
\texttt{e4m3}/\texttt{e5m2})~\cite{ptxisa}. A kernel with $B < 8$ right-hand sides therefore
executes $W_{\mathrm{mma}} = (8/B)\,W$, since the padded lanes are executed whether or not they
carry data. For $B \gg 8$ and for GEMM, $W_{\mathrm{mma}} = W$. At $B{=}1$ this raises the FP8
throughput required to keep up with the memory stream by $8\times$, to approximately $15\%$ of FP8
peak at the full 8~TB/s ($r{=}15$). We include this correction for completeness; the conversion
term of Section~\ref{sec:model} remains the binding one in the analysis below.

\section{Model constants}
\label{sec:constants}

The value of $c_q$ is measured rather than estimated. Converting one element to one modulus on the
cuBLAS int8 emulation path~\cite{schwarz2025ozaki} takes 13 integer instructions. The $3\times$FP8
Karatsuba split~\cite{uchino2026fp8} requires at least 3 additional instructions per modulus per
element to form the FP8 planes. In total, the scaled fp64-to-integer rounding, the per-modulus
reduction, and the FP8 plane conversion cost $c_q = 16$. We use $r = 15$ moduli, which is the
number required to carry the full 53-bit significand, so $c_q r = 240$ integer instructions are
executed per streamed element. For the NVIDIA B300 GPU we use $\Pfp \approx 1.4$~TFLOPS, $\Bmem = 8$~TB/s, and
$\Pint = 75$~TOPS (INT32).

\section{Operational-intensity threshold}
\label{sec:threshold}

Emulation can outperform the native fp64 pipe only if the conversion alone takes less time than
the fp64 computation itself:
\begin{equation}
\frac{W}{\Pfp} \;>\; \frac{c_q r\, Q}{8\,\Pint}
\qquad\Longleftrightarrow\qquad
\OI \;=\; \frac{W}{Q} \;>\; \OI^{*} \;=\; \frac{c_q\, r\, \Pfp}{8\,\Pint}.
\label{eq:threshold}
\end{equation}
On the NVIDIA B300 GPU, $\OI^{*} = (240 \times 1.4)/(8 \times 75) \approx 0.56$~FLOP/B, i.e., 4.5 useful FLOPs
per streamed element. A streaming kernel such as GEMV performs 2 FLOPs per streamed element
($\OI = 0.25$). GEMV is therefore conversion-bound, and its speedup over native fp64 is at most
$0.25/0.56 \approx 0.45\times$, i.e., a slowdown of more than $2\times$, even with $\beta = 1$,
zero-cost MMAs, and full occupancy. In general, every kernel with $\OI < \OI^{*}$ has speedup
$S \leq \varphi^{-1}\,\OI/\OI^{*}$, where $\varphi$ is the fraction of streamed elements that
require conversion. Table~\ref{tab:caps} applies this bound to the kernels in Table~3
of~\cite{matsuoka2026fp8}. We corrected the $\OI$ values for batched GEMV. At the paper's own
$r{=}12$, $\OI^{*}{=}0.45$ and the conclusions are the same.

\begin{table}[ht]
\centering
\begin{tabular}{lccccl}
\toprule
kernel & $\OI$ & $\varphi$ & paper's claim & cap (Eq.~\ref{eq:threshold}) & verdict \\
\midrule
SpMV (CSR/ELL)        & 0.17 & 1          & $1.2\times$  & $0.30\times$ & regression \\
7-pt stencil          & 0.5  & $\sim$0.5  & $3.1\times$  & $1.8\times$  & claim unreachable \\
GEMV ($B{=}1$)        & 0.25 & 1          & ---          & $0.45\times$ & regression \\
batched GEMV, $B{=}2$ & 0.5  & $\sim$1    & $9.2\times$  & $0.89\times$ & regression \\
batched GEMV, $B{=}8$ & 2    & $\sim$1    & $24\times$   & $3.6\times$  & $6.7\times$ below claim \\
dense GEMM            & $\geq 50$ & $\to 0$ & $104\times$ & not binding & unaffected (reuse) \\
\bottomrule
\end{tabular}
\caption{Speedup caps from Eq.~(\ref{eq:threshold}) on the NVIDIA B300 GPU ($r{=}15$, $c_q{=}16$) compared with
the speedups claimed in Table~3 of~\cite{matsuoka2026fp8}. The cap assumes $\beta=1$, zero-cost
MMAs, and full occupancy, and is therefore an upper bound on what emulation can deliver.}
\label{tab:caps}
\end{table}

\section{Precomputed residues}
\label{sec:storing}

One could convert $A$ once, store its residue representation in memory, and remove the conversion
from the inner loop. In that case the kernel streams at least $9r$ bits per element in packed form,
or $3r$ bytes per element as MMA-ready FP8 planes, instead of 8 bytes of fp64. At $r = 15$ this
gives $\beta \geq 9r/64 \approx 2.1$ for the packed form and $\beta = 3r/8 \approx 5.6$ for the
MMA-ready form, and the same cost reappears in the $\beta Q/\Bmem$ term. On-the-fly conversion with
$\beta = 1$ and precomputed residues with $\beta > 2$ are two ways of paying for the same work.
Converting once and reusing the result many times would amortize the cost, but the on-chip
resources needed to hold the converted data are the same resources needed to keep enough memory
requests in flight to reach $\Bmem$. This leaves reuse as an option only for compute-bound GEMM,
where emulation does not need it.

\section{Conditions for invalidating the bound}
\label{sec:falsify}

For GEMV to reach parity with native fp64 on the NVIDIA B300 GPU, Eq.~(\ref{eq:threshold}) requires
$c_q < 8\,\Pint\,\OI/(r\,\Pfp) \approx 7$ instructions per modulus per element, about half the
measured value. The current cuBLAS implementation measures 13 with
INT8, which does not need the 3 or more additional operations per element that the FP8 Karatsuba
method requires. On a GPU with strong fp64, such as B200 with $\Pfp = 40$~TFLOPS, the same
condition is $c_q < 0.25$: converting one element to all 15 moduli would have to take fewer than 4
integer instructions.

\section{Conclusions}

We have shown that the TME model of~\cite{matsuoka2026fp8} omits the per-input residue
deconstruction cost, that this cost is first order for memory-bound kernels, and that on the NVIDIA B300 GPU it
limits Ozaki Scheme~II emulation to below native fp64 performance for GEMV, SpMV, and batched GEMV
with $B \leq 2$, and to well below the claimed speedups for the 7-point stencil and for batched
GEMV with $B = 8$. Dense GEMM is not affected. The bound depends on one measured constant, $c_q$,
and we have stated the value an implementation would have to reach to overturn it.

Finally, we note that there are still many important costs not taken into account with the corrected TME model~\eqref{eq:corrected}, notably the cost of determining scaling factors. This note focuses solely on the costs of quantization and modular reduction.

\end{document}